\documentclass[aps,prb,twocolumn,longbibliography,superscriptaddress]{revtex4-1}

\usepackage{epsfig}
\usepackage{dcolumn}
\usepackage{bm}
\usepackage{amsmath}
\usepackage{amsfonts}
\usepackage{latexsym}
\usepackage{amssymb}
\usepackage{color}
\usepackage{hyperref}
\usepackage[normalem]{ulem}

\newcommand{\tr}{\textcolor{black}}

\usepackage{tikz,xcolor,hyperref}
\definecolor{lime}{HTML}{A6CE39}
\DeclareRobustCommand{\orcidicon}{%
	\begin{tikzpicture}
	\draw[lime, fill=lime] (0,0)
	circle [radius=0.16]
	node[white] {{\fontfamily{qag}\selectfont \tiny ID}};
	\draw[white, fill=white] (-0.0625,0.095)
	circle [radius=0.007];
	\end{tikzpicture}
	\hspace{-2mm}
}

\foreach \x in {A, ..., Z}{%
	\expandafter\xdef\csname orcid\x\endcsname{\noexpand\href{https://orcid.org/\csname orcidauthor\x\endcsname}{\noexpand\orcidicon}}
}

\begin{document}

\title{Real-space analysis of Hatsugai-Kohmoto interaction}

\author{Jan Skolimowski\orcidA}
\email{jskolimowski@magtop.ifpan.edu.pl}
\affiliation{International Research Centre MagTop, Institute of
   Physics, Polish Academy of Sciences,\\ Aleja Lotnik\'ow 32/46,
   PL-02668 Warsaw, Poland}

\begin{abstract}
    The Hatsugai-Kohmoto interaction model has gained a lot of attention in recent years, due to the fact it is exactly solvable in momentum space in any dimension while capturing some key features of the Mott phase. Here a one-dimensional lattice model with this interaction is approached from the real-space perspective, to explore how breaking the translation invariance of a lattice affects the intuition built by studying the exact solution in $k$-space. The ground state properties of chains with periodic and open boundary conditions are calculated and compared with both the exact solution in momentum space, as well as with analogous solutions of the Hubbard model. The results show that introducing hard edges enhances the ferromagnetic correlations and the system undergoes a magnetic transition before reaching the strong coupling limit. Understanding the impact of hard edges is a crucial step toward answering the looming question of the existence of edge states and other topological phenomena in systems with this type of interaction. 
\end{abstract}
\maketitle

\section{Introduction}

Exactly solvable models play an important role in the understanding of physical phenomena. They often provide reference points and intuitions that one can use to understand more accurate, but usually not solvable in general, models. This is the case in strongly correlated electron systems, where the Hubbard model\cite{doi:10.1098/rspa.1963.0204} is the paradigmatic model describing interacting fermions on a lattice. Despite great progress made during its over 50-year-long history in understanding various behaviors observed in this model\cite{doi:10.1146/annurev-conmatphys-031620-102024,d-infty}, a general solution remains elusive. So far, it has only been obtained in one spatial dimension\cite{BetheAnsatz}. Hence there is a need for studies dedicated to exactly solvable\cite{SYK1,baskaran1991exactly}, but often artificial, models of correlated electrons.

One such model, proposed by Hatsugai and Kohmoto \cite{HK}, of infinite-range interaction has been gaining a lot of attention recently. The interest in this Hatsugai-Kohmoto (H-K) interaction was reignited by the observation that it is in the same  high-temperature and strong coupling universality class as the coveted Hubbard model \cite{PhillipsFixedPoint, PhysRevB.108.165136}. Thus, it is capable of reproducing key aspects of the Mott physics. The attractiveness of this model comes also from the fact that it becomes diagonal in momentum space, vastly reducing the complexity of calculating the exact solution. For a single band model it is given by a simple formula\cite{HK}. Hence, in the past four years, multiple studies explored various aspects of model with this interaction, such as Friedel oscillations in a non-Fermi liquid\cite{Zhao_2023}, superconductivity\cite{PhillipsSC}, multi-band physics and its relation to \tr{the} Hubbard model\cite{mai2024new} and the interplay between Mottness and topology\cite{PhillipsTopoMottness}. Few of those papers have also pointed out, that in \tr{the} various models where H-K interaction and topology are intertwined, the system undergoes a topological phase transition without closing the bulk gap in the single particle spectrum\cite{PhysRevB.108.035121,PhysRevB.108.195145,mai2024topological}. This phenomenon originates from the fact that around the \tr{topological phase transition} the system was changing its ground state and a single particle was not capable of capturing the trivialization of the gap through its closure. Hinting that this could be a universal behavior in models with this interaction. This raises a natural question concerning the existence and possible behavior of topological features i.e. edge states\cite{PhysRevLett.95.226801} before the onset of correlation-driven insulating state. A topic that was actively studied for the Hubbard model\cite{SangovanniNature, PhysRevB.98.045133,PhysRevB.90.035116} and topological Kondo insulators\cite{PhysRevB.89.245119}, which also have local Hubbard-type interaction. 

So far, all studies of the H-K model have focused on the solution in the momentum space. Which is the natural basis for this model. In this manuscript, a different approach is taken and the properties of the H-K model are analyzed in real space, which is a more natural basis for the Hubbard-type interaction. This allows one to take the first steps towards understanding the real-space correlation effects produced by the H-K interaction as the system evolves from a metal to an insulator. This should pave the way for future studies on the interplay between topological effects and H-K interaction. 
\tr{In addition,} the large number of real-space studies of the Hubbard model and concepts developed in them\cite{Essler_Frahm_Göhmann_Klümper_Korepin_2005,PhysRevB.50.11403,PhysRevB.51.7934} will serve as a perfect benchmark to explore the similarities between the two models of correlated electrons. 

This work aims to address two main questions: (i) How does the infinite range of the H-K interaction manifest itself in the real-space properties of the finite chain's ground state ?(ii) How is the presence of hard edges, and thus the lack of transnational invariance, altering in the solution? The latter question is crucial, since the exact solution, discussed so far, silently relied on the translation invariance of the system to introduce the momentum-space. In the context of exploring the fate of topological effects, this cannot be done \tr{and} hence \tr{there is} need for a real-space approach. Due to computational complexity, in this manuscript, solely the one-dimensional topologically trivial lattice is concerned. The main finding of this work is that breaking the translation invariance has strong implications for the ground state properties of a chain with H-K interaction. As a function of interaction strength, a chain with hard edges undergoes a transition from a low-spin to a high-spin ground state. This precedes the transition to the strong-coupling limit, in which spatial correlations decay exponentially fast with distance. \tr{This result is independent on the system sizes considered in this work. In the periodic system, the same formation of ferromagnetism is shown to be only a finite size effect. }

\section{Model Hamiltonian}
The H-K interaction\cite{HK} in a particle-hole symmetric form is given by 
\begin{equation}\label{HK_RS}
    \mathcal{H}_{HK}=\frac{U}{2N} \sum_{\sigma,\alpha,\gamma,\Delta}c^\dagger_{\alpha,\sigma}c_{\alpha+\Delta,\sigma}c^\dagger_{\gamma,\bar{\sigma}}c_{\gamma-\Delta,\bar{\sigma}}-\frac{U}{2}\sum_{\alpha,\sigma}n_{\alpha,\sigma},
\end{equation}
where $c^\dagger_{\alpha,\sigma} (c_{\alpha,\sigma})$ is the creation (annihilation) fermionic operator of a particle at site $\alpha$ with spin $\sigma$. In Hamiltonian \ref{HK_RS} (and throughout this text) the \tr{four} Greek letters: $\alpha,\gamma,\Delta,\delta$, will symbolize the site indices and can have values ranging from 0 to $N$, which denotes the number of sites in the system. The summation over $\alpha,\gamma$ is unrestricted over each lattice site, while $\Delta\in [-\alpha,N-\alpha [$. This restricted sum is introduced in such a form to highlight the conservation of the center of the mass of interacting particles. This is the defying property of H-K interaction\cite{HK} in addition to the infinite range. The $\bar{\sigma}$ symbolizes the opposite spin to $\sigma$. The strength of the interaction is controlled by $U$ and does not depend on the distance between the interacting particles. 

In the text, three quantities will be analyzed, as real-space probes of the properties of the system. The first one is the local density of states $A_{\gamma,\sigma}(\omega)$ (LDOS), which provides \tr{information about the spectrum of} the single particle excitations. It is defined as the imaginary part of the retarded local Greens function 
\begin{equation}
    A_{\gamma,\sigma}(\omega)=-\frac{1}{\pi}\Im m G^{R}_{\gamma,\sigma}(\omega),
\end{equation}
where  $G^{R}_{\gamma,\sigma}(\omega)$ \tr{is} the Lehman representation given by
\begin{multline}\label{Lehman}
    G^{R}_{\gamma,\sigma}(\omega)=\\
    \frac{1}{Z}\sum_{n,n^\prime}\frac{\left<n\right|c_{\gamma,\sigma} \left|n^\prime\right>\left<n^\prime\right|c^\dagger_{\gamma,\sigma} \left| n\right>}{\omega+i 0^+ - (E_{n^\prime}-E_n)}\left(e^{-\beta E_n} +e^{-\beta E_{n^\prime}}\right)
\end{multline}
The indices $n,n^\prime$ enumerate the eignestates of a Hamiltonian, $Z$ is the statistical sum and $\beta=1/T$ is the inverse temperature. 
From the exact solution of the H-K model, it is known that in the strong coupling limit, the ground state shows large degeneracy\cite{HK, PhysRevB.108.165136}, which leads to the breaking of the third law of thermodynamics\cite{MelissakiPhD}. To circumvent possibly similar issues in this analysis finite temperature $T=0^+$ will be assumed. \tr{It} is equivalent to calculating the excitations from all possible ground states and averaging the results. A similar route was taken in the original H-K paper\cite{HK}. In the following, only the $\sigma=\uparrow$ channel will be showcased, which due to time-reversal symmetry has to be the same as $\sigma=\downarrow$. Also, the focus of this work is on the bulk ($\gamma=N/2$) and edge ($\gamma=0$) sites. The analysis of the spatial dependence of LDOS is not within the scope of this manuscript. The two other quantities used in this analysis are the equal-time correlation functions: the two-point(2-p) correlator and the 2-p spin-spin correlator. They are  defined, respectively, as 
\begin{equation}\label{crorrelators}
\left<c_{\gamma,\uparrow}c^\dagger_{\delta,\uparrow}\right> \: \mathrm{and } \: \left<\hat{\vec{S}}_{\gamma}\: \cdot \: \hat{\vec{S}}^\dagger_\delta\right>.
\end{equation}
In the latter $\hat{\vec{S}}_\gamma$ is the spin operator at site $\gamma$. As a reference point of the 2-p correlators, a site at one end of a chain $\delta=0$ will be used and the spatial ($\gamma$) dependence of these two correlation functions will be investigated. All correlation functions defined above are given through the expectation value, and the same $T=0^+$ assumption will be used as for the spectral function. All numerical results in this paper were obtained using the exact diagonalization (ED) method. The software used to run the ED calculation was written based on the routines implemented in the {\it libcommute} library.\cite{libcommute}

\begin{figure}
    \centering
    \includegraphics[width=0.5\textwidth]{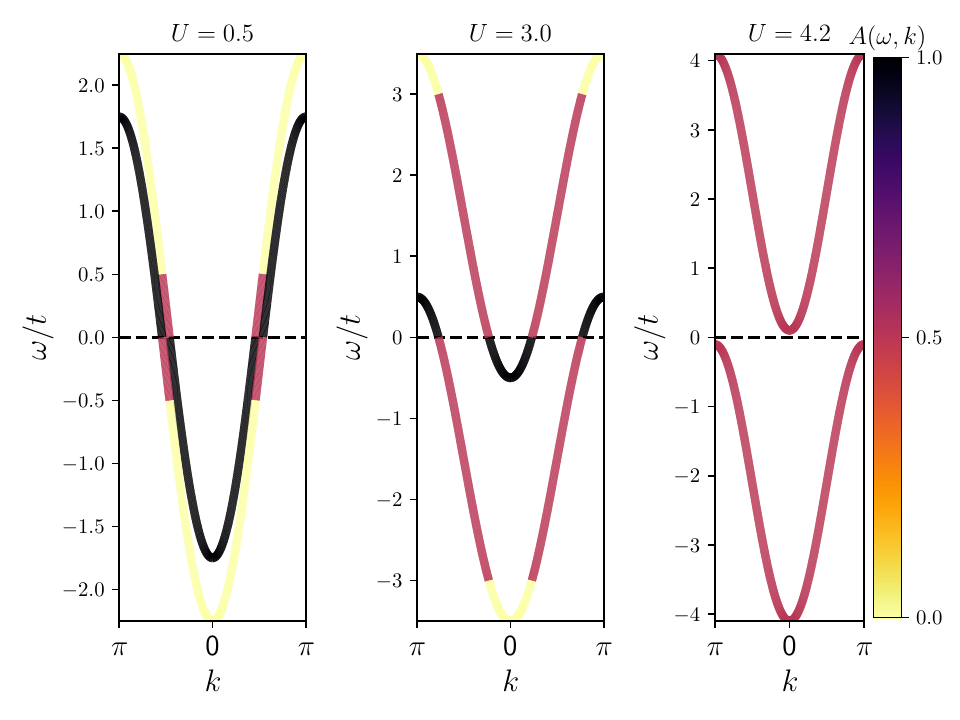}
    \caption{Dispersion relation for a one-dimensional single band Hatsugai-Kohmoto model for three interaction strengths $U$ representing weakly interacting metal (left panel), strongly correlated metal (middle panel) and insulating state(right panel). The colors of the bands represent their spectral weight, which has only three values: zero, one-half, and one.}
    \label{HK-in-k-space}
\end{figure}

As mentioned above, the Hamiltonian \ref{HK_RS} becomes diagonal in the momentum-space and is thus exactly solvable. From this solution, it was shown that the system turns insulating for $U$ equal to the bare bandwidth of the non-interacting electrons\cite{HK}. In the metallic phase, the dispersion possesses discontinuities in the spectral weight, when it jumps from one to one-half and to zero, which is originating from the mixed state ground state with large degeneracy\cite{PhillipsSC}. As illustrated in Fig. \ref{HK-in-k-space} for the half-filled one-dimensional chain, the region with unit spectral weight is centered around the Fermi level and shrinks with $U$ (cf. two leftmost panels). It vanishes at the onset of the insulating phase, where the spectrum consists of two sub-bands of the same bandwidth as the non-interacting electrons, centered at  $\pm\frac{U}{2}$ and with the same spectral weights equal to one-half.

The translation invariance in the system is only possible if the periodic boundary condition (PBC) is assumed. Then each site experience interaction is composed of the same terms. Introduction of hard edges, open boundary condition (OBC), breaks this property because the conservation of the center-of-mass property of the H-K interaction does not allow certain terms to appear in the summation in Hamiltonian \ref{HK_RS}, i.e. the ones for which $\alpha+\Delta$ or $\gamma-\Delta$ falls outside of the system's boundaries. In principle, for the H-K Hamiltonian, the limit of an infinite number of sites does not mean that the edge effect becomes negligible. If one considers a chain of length $N$, then the number of interaction terms that each site of the chain is experiencing with other sites, assuming the PBC, is $N^2$, all with the same strength $U$. As each site index $\gamma$ and $\Delta$ is free to have all $N$ values. Introducing edges means that $\alpha+\Delta$ as well as $\gamma-\Delta$ cannot point outside the $[0,N[$ range. For example, for the middle site ($\alpha=N/2$) $\Delta\in[-N/2,N/2[$ the restriction on the system size does not allow for $\gamma$ to be in its full range. One can count these discarded terms, grouping them by $\Delta$ values. Assuming $\Delta=-N/2$, all $N/2$ terms for which $\gamma\ge N/2$ has to be dropped. For $\Delta=-N/2 +1$  there will be $N/2-1$ terms that give a site index out of bounds. Following the same reasoning until $\Delta=0$, one gets that the number of dropped terms in a finite system, ordered by the value of $\Delta$ is 
\begin{equation}
    \frac{N}{2}+(\frac{N}{2}-1)+\ldots+\left(\frac{N}{2}- (\frac{N}{2}-1)\right)=\frac{N}{4}\left(\frac{N}{2}+1\right)
\end{equation}
\tr{A} similar result can be obtained for $\Delta>0$, with the exception of $\Delta=N/2$, which is outside the allowed range. The $\Delta=0$ case \tr{does not add} any restrictions in the $\gamma$ range. Adding these contributions gives the ratio of dropped terms in the finite system size to the number of terms in a periodic system is
\begin{equation}
   \tr{ \frac{\frac{N}{2}\left(\frac{N}{2}+1\right)-\frac{N}{2}}{N^2}=\frac{1}{4}.}
\end{equation}
This ratio does not go to zero in the limit of infinite system size, and thus the boundary effects for the H-K interaction stay relevant (in terms of the Hamiltonian) even in the thermodynamic limit. For that reason, in the following, the real-space analysis of the H-K interaction for the system with hard edges and with PBC will be conducted in parallel and the difference between the two cases will be highlighted.

\section{Dimer solution}
To gain some first intuitions, a dimer model ($N=2$) with PBC and with OBC (standard dimer) is considered. The periodicity means that an electron on either site of the dimer can hop to its left and right, with the same amplitude. 
It is the simplest non-trivial case, which can be solved analytically in real-space and provide some first insights into the systems dynamic. The solution with PBC should also reproduce some results of the exact solution, as for this case a Fourier transformation can be done, but with $k=0,\pi$ only. The  Hamiltonian \ref{HK_RS} for a dimer with PBC is given by
\begin{multline}\label{dimer_H_HK}
\mathcal{H}_{HK}^{N=2}=\\
=\frac{U}{4}\sum_{\sigma} \left(\sum_{\alpha} \left(2 n_{\alpha,\sigma}n_{\alpha,\bar{\sigma}}-2 n_{\alpha,\sigma}+2 n_{\alpha,\sigma}n_{\bar{\alpha},\bar{\sigma}}\right)
 \right.\\
 \left. 
+ \left(c^\dagger_{0,\sigma}c_{1,\sigma}c^\dagger_{1,\bar{\sigma}}c_{0,\bar{\sigma}}+c^\dagger_{1,\sigma}c_{0,\sigma}c^\dagger_{0,\bar{\sigma}}c_{1,\bar{\sigma}}\right)\right.\\
\left. \left(c^\dagger_{0,\sigma}c_{1,\sigma}c^\dagger_{0,\bar{\sigma}}c_{1,\bar{\sigma}}+c^\dagger_{1,\sigma}c_{0,\sigma}c^\dagger_{1,\bar{\sigma}}c_{0,\bar{\sigma}}\right)\right)
\end{multline}

If the translation invariance is broken, the last line in Eq. \ref{dimer_H_HK} has to be removed. It comes from terms that have $\gamma-\Delta=-1$ and $\gamma-\Delta=2$, respectively, which are only allowed in the modulo system size algebra of the site indices for a periodic system. One can refer to them as "doublon hopping"- because they represent removing a doubly occupied site at one site and creating it on the other.

The particle-hole symmetry of the model means that the ground state will be in the two-electron subspace. The H-K interaction conserves charge and spin, as the Hubbard interaction does, so one can use the same basis states known from the analysis of Hubbard dimer\cite{gebhard1997metal}. These states are  
\begin{align}
\left|1, \pm \right>=\frac{1}{\sqrt{2}}\left(c^\dagger_{0,\uparrow}c^\dagger_{1,\downarrow}\pm c^\dagger_{0,\uparrow}c^\dagger_{1,\downarrow}\right)\left|0\right>, \\
\left|2, \pm \right>=\frac{1}{\sqrt{2}}\left(c^\dagger_{0,\uparrow}c^\dagger_{0,\downarrow}\pm c^\dagger_{1,\uparrow}c^\dagger_{1,\downarrow}\right)\left|0\right>, \\
\left| 3, \sigma\right>= c^\dagger_{0,\sigma}c^\dagger_{1,\sigma}\left|0\right>.
\end{align}

Using the standard ordering of this basis: $\left\{ \left|1, - \right>, \left|2, + \right>,\left|2, - \right> ,\left|1, + \right>, \left| 3, \uparrow\right>,\left| 3,\downarrow\right> \right\}$, which splits the Hamiltonian \ref{dimer_H_HK} into spin-singlet and spin-triplet sub-spaces respectively, the Hamiltonian of a H-K dimer with PBC has the following form 
\begin{equation}\label{H_mat_N_2}
\left[
   \begin{array}{c c c | c c c}
     0 & -4t & 0 &0 &0 & 0 \\
    -4t & 0 & 0 &0  &0 &0\\
    0 & 0 & -U& 0 &0 &0\\
    \hline
    0 & 0 & 0 & -U &0 &0 \\
    0 & 0 & 0 & 0 & -U&0 \\
    0 & 0 & 0 & 0 & 0& -U 
    \end{array}
   \right].
\end{equation}
\tr{The horizontal and vertical lines in the Matrix \ref{H_mat_N_2} and Matrix \ref{H_mat_N_2_dimer} highlight the separation of the Hamiltonian matrix in the two-electron subspace into disjoint blocks consisting of $S=0$ spin singlet states (upper left) and $S=1$ spin triplet states (bottom right).}
From this matrix form, one can see that the periodic H-K dimer undergoes a change in the ground state upon increasing $U$. At weak $U$ it is the bonding state of two spin-singlet states $\left|1,- \right>$ and $\left|2,+ \right>$, with energy $E=-4t$. The same is true for the Hubbard dimer with PBC. The difference between the two is, that Hubbard interaction favors the contribution from $\left|1, - \right>$, while H-K interaction does not. At $U=4t=U_c$ all states, apart from the anti-bonding state in the spin-singlet subspace become degenerate. The five-fold degeneracy is later reduced to a four-fold degeneracy for $U> U_c$, with all three spin-triplet states and one spin-singlet state ($\left|2, - \right>$) becoming the ground states. This is in stark contrast to the Hubbard model, where the ground state is never degenerate and always has $S=0$. Here, the H-K interaction clearly promotes ferromagnetic alignment of the two spins. \tr{The ground state subspace is not only spanned by the $S=1$ states, whose degeneracy is a consequence of the conservation of the $S_z$ component of the spin by the Hamiltonian \ref{dimer_H_HK}, but there is also a contribution from a state with  $S=0$.} A similar conclusion, that H-K shows a ferromagnetic instability, was reported for the QSH systems\cite{PhillipsTopoMottness} and for $k$-dependent interaction strength version of the H-K model\cite{PhysRevB.103.024529}. This analysis shows that it can be a generic feature of this interaction in finite systems. \tr{This} is even enhanced by the breaking of the translation symmetry, as will be shown now.

The H-K dimer (with OBC) Hamiltonian, in the same basis as previously, has the following form 
\begin{equation}\label{H_mat_N_2_dimer}
\left[
   \begin{array}{c c c | c c c}
     0 & -2t & 0 &0 &0 & 0 \\
    -2t & -\frac{U}{2} & 0 &0  &0 &0\\
    0 & 0 & -\frac{U}{2}& 0 &0 &0\\
    \hline
    0 & 0 & 0 & -U &0 &0 \\
    0 & 0 & 0 & 0 & -U&0 \\
    0 & 0 & 0 & 0 & 0& -U 
    \end{array}
   \right].
\end{equation}
On the level of the \tr{form} of the Hamiltonian, the difference to the periodic analog is the reduction of the kinetic energy ($-4t\rightarrow -2t$), due to reduced number of hoppings, and the lack of the "doublon hopping" term. This later term was lowering the energy of the $\left| 2,\pm\right>$ states, with doubly occupied sites. Removing it turns out to be sufficient to give rise to a clear magnetic transition in the system. From the Hamiltonian 
Matrix \ref{H_mat_N_2_dimer} one can see, that at large $U$ the ground state is only within the triplet subspace with three-fold degenerate energy $E_{GS}=-U$ \tr{ in the bottom right corner}. The degeneracy with the spin-singlet state, present in the periodic model, is lifted. On the other hand, at small $U$ the ground state is only within the spin-singlet subspace. As a consequence, in a dimer as $U$ decreases the ground state jumps from (just) the spin-triplet subspace to (just) the singlet subspace, where it has the energy
\begin{equation}
    E_{GS}=-\sqrt{(2t)^2+\left(\frac{U}{4}\right)^2} -\frac{U}{4}
\end{equation}
Through comparison of the energies of Hamiltonian matrix \ref{H_mat_N_2_dimer}, one obtains the critical interaction strength $U=U_c=2\sqrt{2}t\approx2.83 t$ for the transition between different magnetic states. This result is below the critical $U$ of the periodic case. In addition, although the ground state in the small-$U$ limit consists of the same states as in both the H-K dimer with PBC and the Hubbard dimer the weight of each contributing state is different for all three cases. The lack of the "doublon hopping" results in the ground state for $U<U_c$, which favors the $\left|2, + \right>$ state, as opposed to equal contribution in the bonding state, discussed previously. A similar comparison between a dimer with PBC and OBC in the case of the Hubbard interaction, does not show any qualitative changes upon changing the boundary condtions. It is understandable, as it is purely local interaction, unlike the H-K interaction which can have infinite range. 
The root of the spin alignment in a dimer for large $U$ in the H-K model can be traced back to the last density-density term in the second line of Hamiltonian \ref{dimer_H_HK}. This term is not the usual density-density term as in e.g. the extended Hubbard model\cite{Lin1995}, where large densities in the adjacent sites are penalized. Here the energetically unfavorable situation is when spins at the neighboring sites anti-aligned. Hence the high-$U$ state prefers their ferromagnetic alignment, instead of anti-ferromagnetic as in the Hubbard model. 
This effect is even more amplified if one omits all terms in the last two lines of the Hamiltonian \ref{dimer_H_HK} altogether, and keeps only the density-density terms. Then, the large-$U$ ground state is only doubly degenerate and belongs to the subspace of $\left| 3, \sigma\right> $ states, irrespective of the boundary conditions.

To analyze the spectral properties one has to also look at the Hamiltonian matrices in the one- and three-electron subspace. The p-h symmetry means they will have the same form with doubly degenerate states with energies 
\begin{equation}
E^{1,3}_{OBC,\pm}=-\frac{U}{2}\pm t
\end{equation}
for the dimer with OBC and 
\begin{equation}
E^{1,3}_{PBC,\pm}=-\frac{U}{2}\pm 2t
\end{equation}
for the periodic case. As these states correspond to the effectively non-interacting model, these energies and their corresponding eigenstates are the same as in the Hubbard model for a dimer with OBC and PBC, respectively.

From this one can calculate the location of poles in the single particle spectrum from Eq. \ref{Lehman}. For simplicity, the focus will be on the particle-like excitations, from the two-electron to three-electron subspace. The p-h symmetry secures, that the hole-like excitations will have exactly opposite energies but the same spectral weights. 
For a dimer with PBC below the critical $U$
\begin{equation}\label{HK_low_U_exct_PBC}
\Delta_\pm^{HK-PBC} = E^{3}_\pm -E_{GS}=-\frac{U}{2}\pm 2t +4t 
\end{equation}
and for $U>U_c$
\begin{equation}\label{HK_high_U_exct_PBC}
\Delta_\pm^{HK-PBC}  = E^{3}_\pm -(-U)=\frac{U}{2}\pm 2t .   
\end{equation}
Similarly, for a dimer with OBC below the critical $U$
\begin{equation}\label{HK_low_U_exct}
\Delta_\pm^{HK-OBC} =\sqrt{(2t)^2+\left(\frac{U}{4}\right)^2} -\frac{U}{4}\pm t 
\end{equation}
and for $U>U_c$ the single particle excitation will have energies  
\begin{equation}\label{HK_high_U_exct}
 \Delta_\pm^{HK-OBC}  = E^{3}_\pm -(-U)=\frac{U}{2}\pm t .   
\end{equation}
For comparison, the Greens function of a Hubbard dimer has poles at 
\begin{equation}\label{Hub_exct}
    \Delta_\pm^{Hub-OBC}=\sqrt{(2t)^2+\left(\frac{U}{2}\right)^2} \pm t.
\end{equation}
and assuming PBC in a Hubbard dimer gives 
\begin{equation}\label{Hub_exct_PBC}
    \Delta_\pm^{Hub-PBC}=\sqrt{(4t)^2+\left(\frac{U}{2}\right)^2} \pm 2t.
\end{equation}

From Eq.\ref{Hub_exct} and Eq.\ref{Hub_exct_PBC} one can see that the poles of the Hubbard dimer Greens function, independently of boundary conditions, start to shift away from the Fermi level as soon as the interactions are switched on. 
For the H-K dimer initially ($U< 2t$) the excitations start to approach the Fermi level (cf. Eq. \ref{HK_low_U_exct}) nearly linearly with $U$. Only after the ground state jumps to the triplet subspace, and its energy becomes $-U$, the excitations start to gain energy and drift away from the Fermi level.

\begin{figure}[h!]
\flushleft
\includegraphics[width=0.48\textwidth]{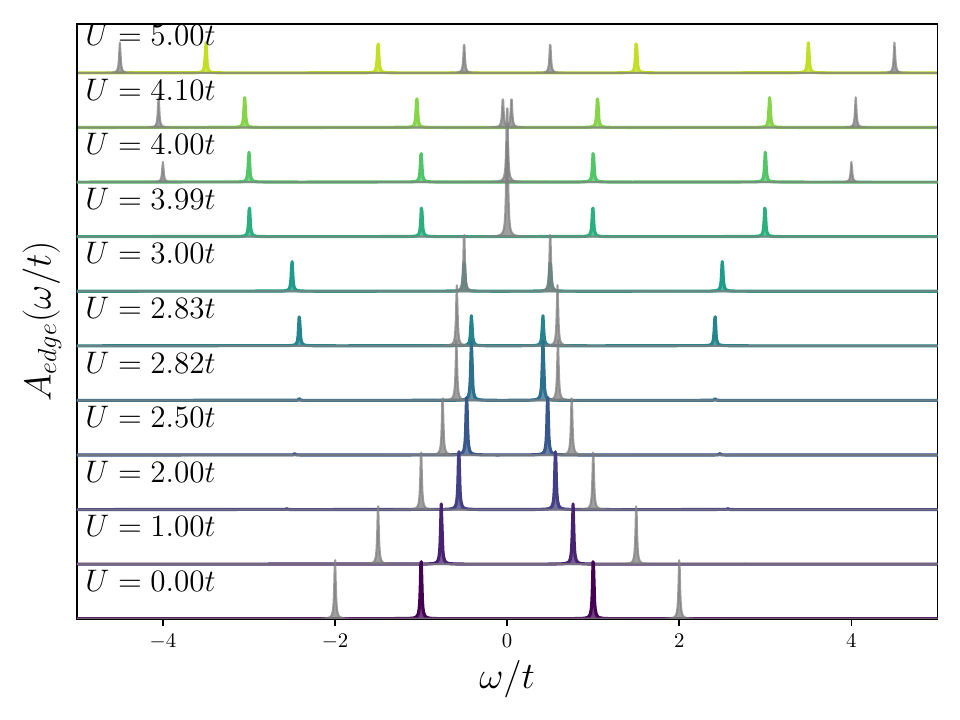}
\caption{Comparison of the LDOS of the H-K dimer with OBC (color) and with PBC (grey) for various interaction strengths across the transition from the weak($U=0$) to strong coupling limit ($U=5t$). Both models undergo a change in the ground state as a function of $U$. For the periodic dimer $U_c=4t$ and the dimer with OBC has the transition at $U\approx 2.83t$. In both cases, the transition is accompanied by a rapid change in the spectrum.}
\label{HK_vs_PBC_dimer}
\end{figure}
To illustrate the above-discussed spectra of the H-K dimer with  OBC and PBC, in Fig. \ref{HK_vs_PBC_dimer} the two are put on top of each other for various $U$ values ranging from weak to strong coupling. In color are the spectra of the OBC case and in grey are the results assuming PBC. At $U=0$ (bottom panel) the two models have peaks at different locations due to different kinetic energy. For a dimer assuming periodicity effectively just doubles the hopping amplitude. Until $U<2.28t$ the qualitative behavior of the two models is similar, with peaks approaching the Fermi level and following the Eq. \ref{HK_low_U_exct} and Eq.\ref{HK_low_U_exct_PBC}. What is different is that the dimer with OBC has a small but finite spectral weight in the secondary peaks with energies higher by $2t$ than the main. The analogous peaks in the periodic model do not carry any spectral weight. This is the same feature that the infinite (periodic) system solution has. To see that, one has to consider the Fourier transformation of the operators used to calculate the LDOS shown in Fig. \ref{HK_vs_PBC_dimer}. They have contributions from two momenta $0,\pi$. Since the two momentum blocks are disconnected for the H-K interaction the local density of states is just a simple sum of the two contributions from the $k$-resolved spectrum. From Fig. \ref{HK-in-k-space} one can see that at these $k$-values only one branch of the dispersion contributes and it has unit weight. The other has exactly zero weight and is not visible. Which is what was observed in the periodic dimer solution. Following this relation with the continuous $k$ solution, after the transition to an insulating phase ($U>4t$)the spectral weight is redistributed equally between the two branches at $k=0$ and $k=\pi$. This is also seen in the periodic dimer solution, where for $U>4t$ the height of the peaks drops by half. At exactly $U=4t$ the middle peak is a sum of contributions from two sub-bands, thus its weight is not halved. The transition from a non-degenerate to a degenerate ground state is nonetheless already visible in LDOS in the form of suddenly appearing high energy excitations at $\omega=\pm 4t$, with half of the spectral weight as the main peaks for $U<U_c$. Such a discontinuous evolution of the excitation spectra would not be possible if the ground state stayed non-degenerate. In addition, the four-fold degeneracy of the ground state of the dimer with PBC in large-$U$ limit is matching the $2^N$ spin-degeneracy of the $N$-electron ground state of the exact solution\cite{MelissakiPhD}. This connection between the dimer with PBC and the $k$-resolved solution shows the relevance of the former for analysis of the opening of the gap in the one-dimensional chain, due to the opening happening at $k=0$ and $k=\pi$. Another picture emerging from the analysis of a dimer is that finite system size can reduce the ground state degeneracy in the correlated metal phase, without any additional terms in the Hamiltonian. The degeneracy in this interaction strength regime comes from the spin degree of freedom of the singly occupied $k$-states. Since the finite system size controls the allowed $k$-values it can be tuned to remove all the singly occupied $k$-states from the system. This is the case for the dimer with PBC, as shown above. The property of tuning the ground state degeneracy of local in $k$ interaction could be of interest for the cold atom community, where similar systems could be realized\cite{PhysRevLett.120.040407, PhysRevLett.127.130401}. 

As shown in Fig.\ref{HK_vs_PBC_dimer}, the dimer with OBC undergoes a magnetic transition at $U\approx2.83t$. Similarly to the periodic case, the half of spectral weight suddenly jumps to higher energy poles as $U$ crosses the $U_c$. From this point on the local density of states evolves monotonically with $U$ and pairs of excitations above and below the Fermi level are drifting apart. The excitations at large $U$  with OBC and with PBC behave qualitatively the same, but they are located at slightly different positions, due to the difference in the kinetic energy term. 
\begin{figure}[h!]
\flushleft
\includegraphics[width=0.48\textwidth]{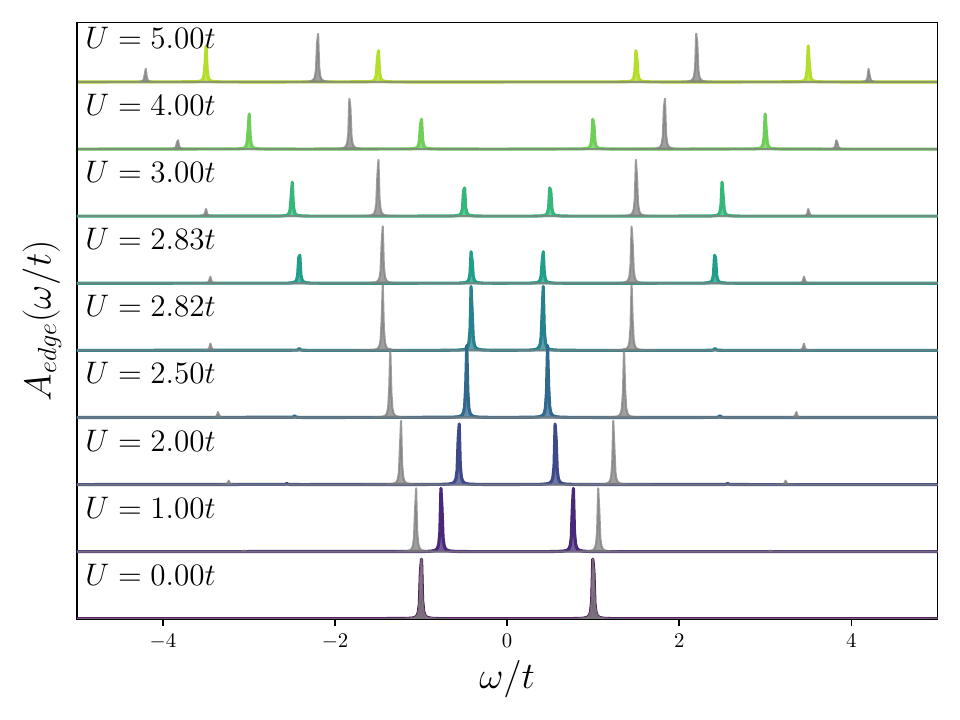}
\caption{Comparison of the LDOS of the H-K dimer with OBC (color) and the corresponding Hubbard dimer (grey) for various interaction strengths $U$. The former shows a rapid change around $U\approx 2.83t$, where the ground state moves from the spin-singlet to the spin-triplet subspace.}
\label{HK_vs_Hub_dimer}
\end{figure}
Figure \ref{HK_vs_Hub_dimer} shows how the same spectra of a dimer with OBC compare to the analogous Hubbard dimer. The biggest difference is naturally for $U<2.83t$. The attraction between the excitations in H-K model is in contrast to the repulsion of the analogous excitations in the Hubbard model. After, the magnetic transition in the H-K model the two solutions behave qualitatively the same. Following Eq.\ref{HK_high_U_exct} and Eq. \ref{Hub_exct}, one can see that in the limit of $U \gg t$ the two spectra approach each other deep in the strong coupling limit. What is different between the two, is that the ground state favors different spin alignments. The natural question is, whether this magnetic instability of the high-$U$ phase of the H-K model is enhanced by the extremely small system size of a dimer or it will survive in larger systems as well. If so, what is its spatial dependence?
\tr{
\section{Finite size scaling of spin-spin correlator}
}
\tr{
As the system size increases beyond $N=2$ providing an analytical evidence for ferromagnetic correlations in H-K model becomes not feasible. To overcome this problem one can do a finite size scaling analysis of a the 2-p spin-spin correlator, defined in \ref{crorrelators}, between the neighbouring sites, which is obtained using the exact diagonalization. }
\begin{figure}
    \flushleft
    \includegraphics[width=0.5\textwidth]{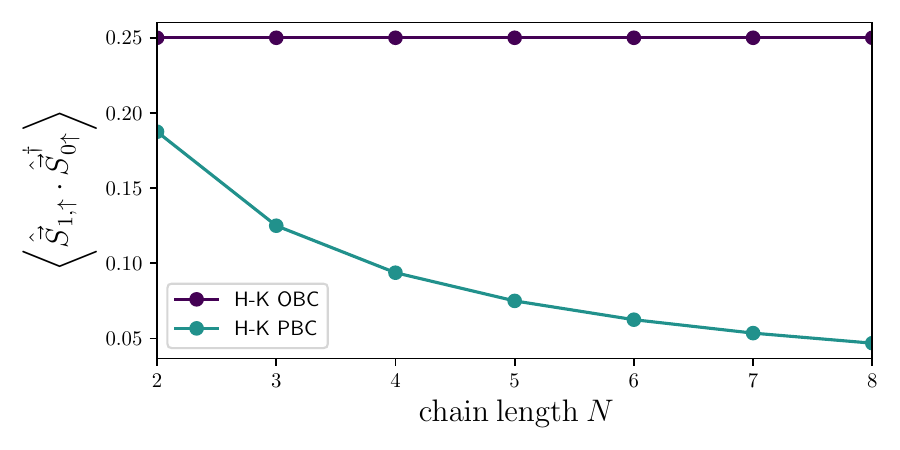}
    \caption{Finite size scaling of the two-point spin-spin correlator between the nearest neighboring sites as a function of system size $N$ for the H-K chain with OBC and PBC. All results are for $U=5$ which is large enough to secures that the system is in the high-$U$ solution.}
    \label{HK_OBC_vs_PBC_decay}
\end{figure}
\tr{
Results comparing the H-K models with OBC and PBC for the high-$U$ state ($U=5$) are displayed in Fig. \ref{HK_OBC_vs_PBC_decay}. They show that the ferromagnetic correlations, non-zero for a dimer, in a system with OBC are unaffected by the increase of the system size $N$. Exactly the same value of $0.25$ was obtained for all $N$ studied, and thus its likely that it should survive the limit of $N\rightarrow\infty$. On the other hand, the spin-spin correlations for the periodic model do show a rapid decay. Already at $N=8$ they are almost order of magnitude smaller then for $N=2$. This trend suggests that the ferromagnetic correlations in the high-$U$ phase could be strongly damped if not completely destroyed by the thermodynamic limit. Nonetheless, their relatively large non-zero value in the periodic dimer analysis was a finite size effect. The discrepancy in the finite size scaling between the H-K model with OBC and PBC illustrates how sensitive the model is to the boundary conditions. Their role should not be neglected in future studies, as they leads to distinct and measurable change in the properties of the system.
}\tr{
Because the ferromagnetic correlations of the high-$U$ phase of the system with OBC showed no dependence on $N$, in the next section the results for the chain with $N=8$ will be analyzed in more details to unveil the spatial dependence of the 2-p correlators and the impact of increased number of single particle excitations on LDOS. Since the goal of this work is to highlight the qualitative differences in the H-K model with hard edges the exact value of $N$ is less important. The results from Fig. \ref{HK_OBC_vs_PBC_decay} show that the long-range nature of the H-K interaction cancels even (eventual) odd-even effects that could impact the high-$U$ limit solution in small systems. On the other hand, since the asymptotic value of the spin-spin correlations in $N\rightarrow\infty$ of a periodic chain is not possible to determine from this study, the $N=8$ represents the case where these correlations are still present yet are much smaller then in a similar system with OBC.  
}
\section{The chain with $N=8$ sites}

\begin{figure}
    \flushleft
    \includegraphics[width=0.5\textwidth]{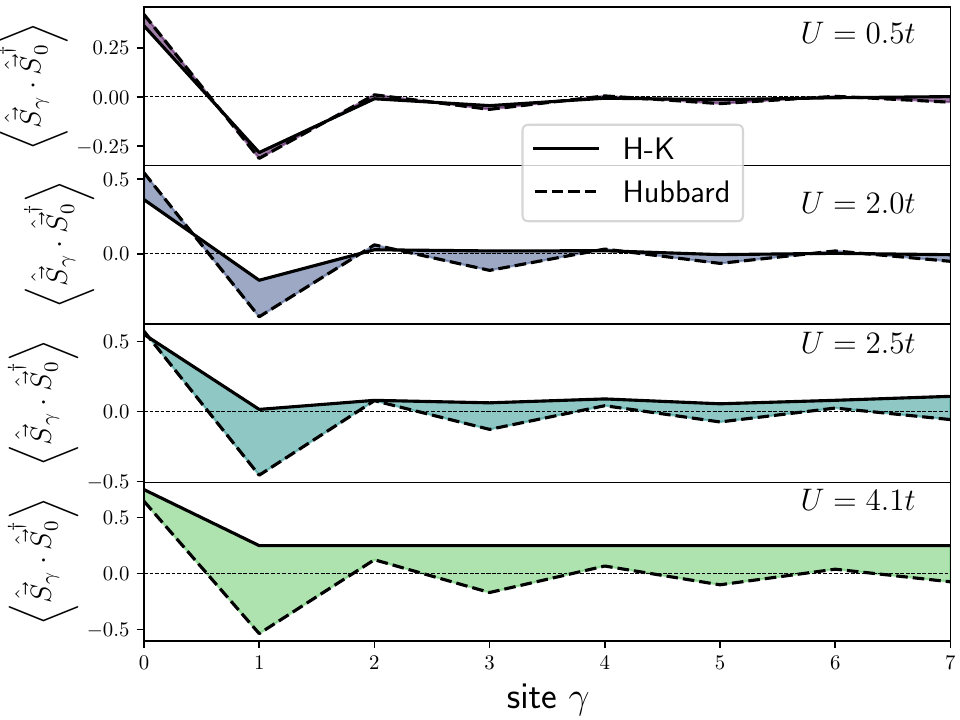}
    \caption{Two-point spin-spin correlator for a chain with OBC of $N=8$ sites with H-K (solid line) and Hubbard (dashed line) interactions, for various interactions strengths $U$. The area between the two curves is colored to highlight the differences between them. The $U$ values show the transition of the spin-spin correlations in the H-K model from anti-ferromagnetic(top panel) to ferromagnetic (bottom panel).}
    \label{HK_vs_Hubbard_ss_N_8_bulk}
\end{figure}

Figure \ref{HK_vs_Hubbard_ss_N_8_bulk} displays the spatial-dependenc of the 2-p spin-spin correlator for various interaction strengths from weak to strong couplings for a chain with OBC. The Hubbard model at half-filling is used as a reference, since it is famously having short-range anti-ferromagnetic correlations\cite{PhysRevLett.64.2831,PhysRevB.42.10553,PhysRevB.50.11403}. 

The solid lines are the results for a chain with H-K interaction and the dashed lines are the result for the analogous Hubbard model. The colored region between the curves is added to highlight the discrepancies between the results for the two models. The Hubbard model results show an oscillatory behavior around zero as a function of distance from the $0$-th site. The alternating sign of the spin-spin correlator, with negative values at odd distances and positive and even distances ($\gamma$) from the reference spin, signals the anti-ferromagnetic nature of the coupling between the spins at adjacent sites. The amplitude slightly increases with the interacting strength $U$, reflecting the enhanced correlations, and decays with distance showing the lack of long-range correlations in the spin-spin channel of the Hubbard model.
Similar behavior is also present in the H-K model, but only up to a certain small $U$ value. In agreement with the dimer analysis, where the ground state did hold a resemblance to the Hubbard model as far as spin-degree of freedom is concerned. Between $U=2t$ and $U=2.5t$, two middle panels of Fig. \ref{HK_vs_Hubbard_ss_N_8_bulk}, the anti-ferromagnetic correlations give way to the ferromagnetism. The 2-p spin-spin correlator for $U=2.5$ is already \tr{showing} only positive values independently of the distance $\gamma$. Initially, small oscillations of the spin-spin correlator are observed in the H-K model at $U=2.5$. But, as shown in the bottom panel of Fig.\ref{HK_vs_Hubbard_ss_N_8_bulk}, for $U=4$ these correlations do not decay with $\gamma$ anymore. Instead, they saturate at one-quarter for all $\gamma>0$. \tr{This lack of any decay with $\gamma$ of ferromagnetic correlations show that it is not just an edge effect but also appears in bulk}. This supports the idea that the ground state of a H-K model with OBC in the high-$U$ limit is neither an anti-ferromagnetic as it would be in the case of the Hubbard model\cite{PhysRevLett.62.1201,PhysRevB.51.7934} nor a non-magnetic as the momentum-space solution would suggest\cite{CONTINENTINO1994619}. Instead it is a ferromagnet, with non-vanishing long-range positive spin-spin correlations. 

\tr{Entering the high-$U$ phase is also signalled by raise of the degeneracy of the ground-state. Interestingly, the high-$U$ ($U>4$) ground state of a $N=8$ chain with PBC is exactly $256=2^8$ times degenerate, which is the behavior reported in the dimer and the feature of the exact continuous $k$ solution. On the other hand, the ground state of the system with open boundaries do not approach the degeneracy that exponentially depends on the number of electrons. This difference combined with the analogous behaviors in the finite periodic system and the continuous $k$ solution, leads to believe that the presence of boundaries effectively introduces an interaction between the spins in the momentum space. As a result the magnetism in finite H-K model appears naturally, in contrast to the $k$-space solution, in which one had to add artificial terms to the Hamiltonian that favored one of the spin orientations, and thus removed the huge degeneracy of the exact (periodic) solution\cite{MelissakiPhD,PhysRevB.103.024529,PhysRevB.108.165136}. }

\begin{figure}
    \flushleft
    \includegraphics[width=0.5\textwidth]{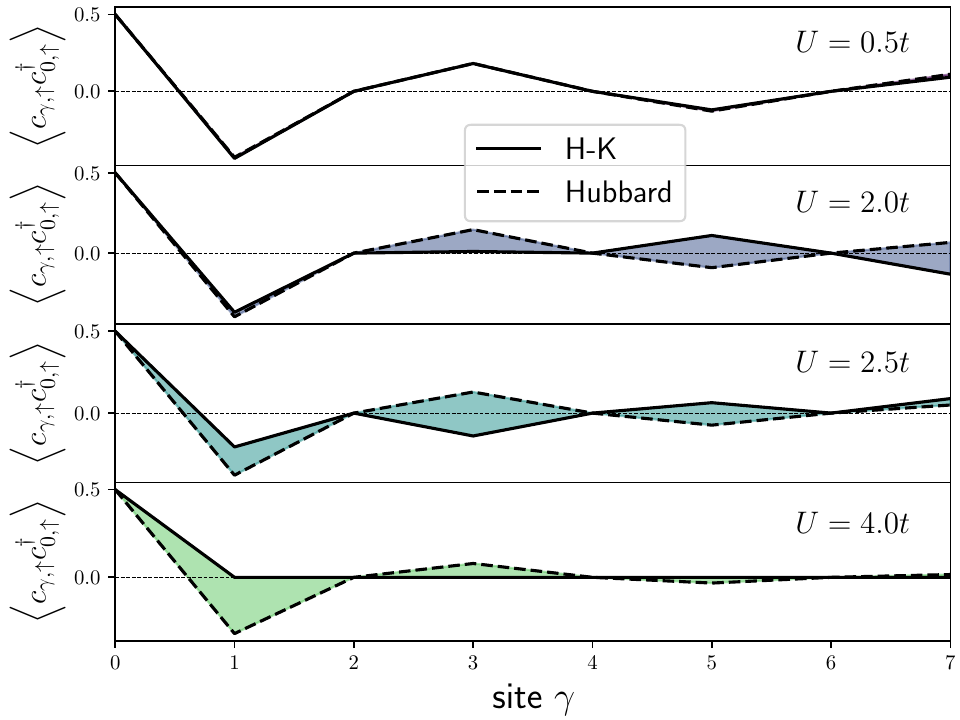}
    \caption{Two-point correlator for a chain with OBC, $N=8$ sites with H-K (solid line) and Hubbard (dashed line) interactions, for various interaction strengths $U$. The region between the two curves is colored to highlight the differences between them.}
    \label{HK_vs_Hubbard_2p_N_8_bulk}
\end{figure}
For completeness, the 2-p correlator is displayed in Fig.\ref{HK_vs_Hubbard_2p_N_8_bulk}, which was recently reported\cite{Zhao_2023} to show a surprisingly quick decay with distance in the high-$U$ solution of the H-K model, which could be used as criterion for the strong-coupling limit. 
The results presented here for the 2-p correlation reproduce this behavior, cf. bottom panel. The comparison with the Hubbard model shows, that this rapid decay is even faster for the long-range H-K interaction than for the purely local Hubbard interaction. \tr{Despite the fact, that the contribution of the Hubbard terms in the H-K interaction decays with $N$, cf. Eq. \ref{HK_RS}.} For $N=8$ sites the Hubbard model, even though it has strong local terms that favor charge localization, has correlations that can spread even to the other end of the chain. 
\tr{This decay is not directly connected to the formation of ferromagnetism in the H-K model with OBC. Comparing results for all four values of $U$ displayed in Fig.\ref{HK_vs_Hubbard_2p_N_8_bulk}, the damping of the spatial dependence of 2-p correlator with $U$ does not match the magnetic transition shown in Fig. \ref{HK_vs_Hubbard_ss_N_8_bulk}. In the two middle panels, for $U$ values around the magnetic transition point, the 2-p correlator still extends across the whole chain. Nonetheless, the role of the tendency of the H-K interaction to promote high spin state cannot be overlooked in the damping of 2-p correlator. The quantity, which describes moving one electron from one site to another, in a high spin state has strong restrictions due to Pauli principle. At small-$U$ the close resemblance between the low-$U$ ground states of Hubbard and H-K models is also reflected in the 2-p correlator. In the top panel, the two models produce almost identical behaviors.
}

\begin{figure}
    \flushleft
    \includegraphics[width=0.5\textwidth]{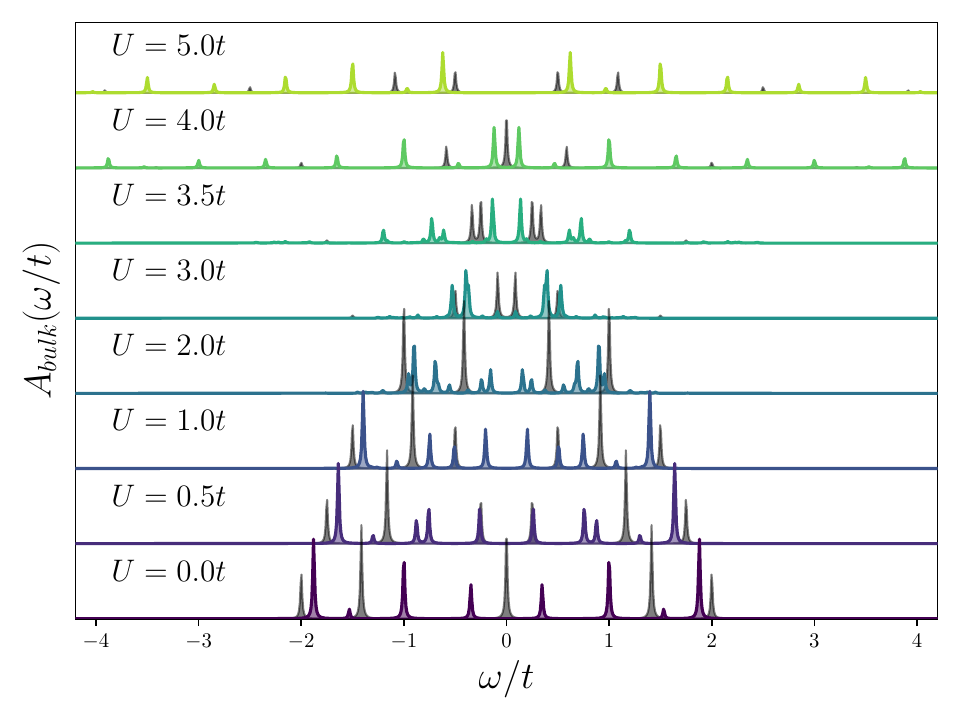}
    \caption{Comparison of local density of states at the middle of $N=8$ site chain($\gamma=4$), referred to in the text as bulk, for various interaction strength $U$. The grey peaks represent the poles of the Greens function of the chain with H-K interaction and PBC. The colored peaks are the excisions of a system with OBC. The interaction strength spans from $U=0$ (non-interacting)  to $U=5t$ (strong coupling limit).}
    \label{HK_finite_vs_PBC_N_8_bulk}
\end{figure}
Lastly, the spectral properties of a larger chain of length $N=8$ are compared, similarly as it was done for the dimer case. In the Fig. \ref{HK_finite_vs_PBC_N_8_bulk} local density of states at the bulk ($\gamma=4$) of a chain with OBC (color) with its periodic analog (grey) are contrasted. The figure shows a qualitatively similar evolution of LDOS with $U$ as observed for the dimer. At smaller $U$ values, the overall bandwidth (distance between outermost peaks) of the excitations shrinks with the increase of the interaction strength, cf. two bottom panels of Fig.\ref{HK_finite_vs_PBC_N_8_bulk}. At larger $U$ this trend is reversed as peaks with very small spectral weight start to appear at higher energies. As a result, the spectral weight becomes more spread across a larger energy range. For the solution with PBC, this could be again understood in terms of the exact solution with the continuous momentum $k$. For $N=8$ the local operator, used in calculating the LDOS, has a contribution from eight discrete values of $k$ ( $k=\frac{2\pi}{8}\times n$ with $n\in[0,8[$ ), all with an appropriate phase factor. 
This discretization of $k$ is the reason why the $U=0$ solution with PBC (grey peaks in the bottom panel of Fig.\ref{HK_finite_vs_PBC_N_8_bulk}) does not look like the LDOS of an infinite chain. The $k$-values creating the outermost peaks ($k=0,k=\pi$) are taken only once, whereas the peaks inside the band are taken twice e.g. $k=\frac{\pi}{4}$ and $\frac{7\pi}{4}$. Hence, the peaks at the edges carry half of the weight of the peaks around the Fermi level. At small $U$ most contributions come from the part of the band with unit spectral weight. As it shrinks with $U$ the distance between the outermost peaks of LDOS is also reducing, cf. four bottom panels in Fig. \ref{HK_finite_vs_PBC_N_8_bulk}.  At certain $U$ the discrete $k$ values start to capture the parts of the bands with one-half spectrum. As a result, the excitations with smaller weights but at higher energy start to appear. This breaks the trend of narrowing the support of the LDOS. The spectral weight carried by them is not simply either one-half or one anymore. It is changed by the interference between the spectra of different $k$-points. For $N=8$ the Fourier transformation of an operator creating an electron at a given site has also contributions from operators with complex amplitude. This allows for interference between the spectrum at different $k$. A similar redistribution of spectral weight from just a few main peaks to much smaller peaks spread over a larger energy range is observed in the system with OBC, already at $U=2$. The lack of periodicity does not allow to simply interpret it in terms of the exact solution in $k$. Nonetheless, the similarities between the solutions of the chain with PBC and OBC lead to believe the key aspects of the exact solution are present in both cases. 
Above $U=4t$ the $k$-space solution has two separate bands with a gap of $U-4t$ at $k=0$. These lower band edges for $U=5t$ should appear at $\omega=\pm 0.5 t$ and the PBC model captures it, as shown in the top panel of Fig. \ref{HK_finite_vs_PBC_N_8_bulk}. The position of the outermost edges is also recovered. 
To limit the computational time in case of the huge degeneracy of the ground state of the system with PBC in the large $U$ limit($U\ge 4$), the calculation of the spectral function was limited to up to 60 lowest energy excited states within each invariant subspace of the Hamiltonian. This number was established through scaling analysis of the LDOS as the lower bound on the number of states kept do not produce any significant changes to the final result.

\begin{figure}
    \flushleft
    \includegraphics[width=0.5\textwidth]{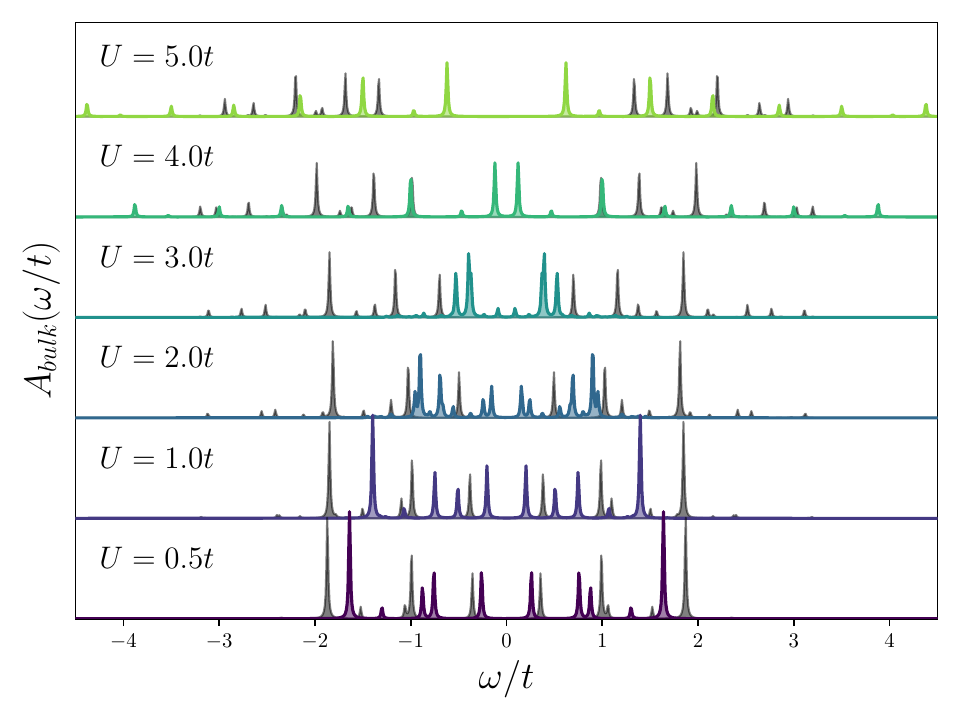}
    \caption{The bulk LDOS (in the middle) of a chain with OBC of length $N=8$ with H-K interaction(color) and Hubbard interaction (grey) for various strengths $U$ from weakly to strongly interacting limit.}
    \label{HK_vs_Hubbard_N_8_bulk}
\end{figure}

Figure \ref{HK_vs_Hubbard_N_8_bulk} shows the local density of states in the middle of a one-dimensional chain with OBC and H-K interaction (color) and compares it against the solution of an analogous Hubbard model (grey). 
Once again one can see in this comparison a resemblance to the results obtained in the analysis of a dimer. Namely, the lowest laying poles of the H-K local Greens function attract, while the Hubbard model shows an opposite trend. The same trend is reflected by the strong peaks (with large spectral weight) at higher energies of the H-K LDOS and is reversed only around $U\approx 4t$. Another difference between the two models is seen in the formation of the small peaks at higher energies. For the Hubbard model, these start to form outside the $U=0$ band edges already at small $U$, cf. $|\omega/t|>2$. As the interaction becomes stronger these accumulate more weight and spread over a larger energy range. For the H-K model, these features form at a smaller rate with $U$ and mostly within the initial energy range of the $U=0$ LDOS ($\omega\in[-2t,2t]$). They become visible for $U\approx 2t$ and higher. Assuming that the similarity to the exact solution holds in the system with OBC, the largest peaks in the LDOS of the H-K chain would follow the band edges of the unit spectral weight part of the dispersion, and the small peaks outside of them come from the band with one-half weight.

\section{Summary}
This manuscript was dedicated to the analysis of the real-space properties of the ground state of a chain with Hastugai-Kohmoto interaction, which by definition has infinite range and conserves the center-of-mass of the interacting particles. It was shown that the form of this interaction is sensitive to the presence of boundaries irrespective of the size of the system, due to the latter property. By analyzing first the solution of a dimer with open and periodic boundary conditions, it was shown that the system can have a transition from a spin-singlet to a spin-triplet ground state, as a function of interaction strength $U$ in the case of OBC. The favoring of $S=1$ ground state stems from the unusual density-density term in the H-K interaction, which penalizes the anti-alignment of the neighboring spins. This interaction is partially screened by the "doublon hopping" term, present in the case of PBC. The ferromagnetic spin correlations are also present in the periodic dimer, but are hindered by the degeneracy between one spin-singlet state and the $S=1$ states. This is in stark contrast to the Hubbard model, where the anti-ferromagnetic correlations in the $S=0$ ground state are promoted at all interaction strengths $U$.
\tr{In order to rule out finite size effects as main cause of this observation a scaling analysis of spin-spin correlator between the neighbouring sites was made for a system with open and periodic boundary conditions. It showed that the system with hard edges indeed promotes a high spin ground state, as this function showed no change with the system size. In contrast, the ground state of a periodic chain with H-K interaction had this correlation function decaying with increasing number of sites. To illustrate the effects stemming from increased systems size, a more in-depth analysis of a chain with $N=8$ sites was made. The 2-p spin-spin correlator for $N=8$ confirmed the presence of ferromagnetic correlations even at larger distances from the edge and also for moderate interaction strengths $U<4t$. The emergence of a ferromagnetic correlation did not come with an immediate decay (with distance) of the two-point correlator. This property of the strong coupling solution of the H-K model happens at higher interaction strength. Unexpectedly, a comparison with the analogous Hubbard model showed that the 2-p correlator decays much faster in the H-K model for the same $U$. This means that correlations in an (in principle) infinitely long interacting model decay faster than in the case of purely local interactions. Lastly, analysis of the spectral functions showed the same narrowing of the spectra upon increasing $U$ for either boundary condition and the resemblance of PBC the exact $k$-space solution was further confirmed.}

\begin{acknowledgments}
The author would like to thank W. Brzezicki, M. Wysokiński, C. Mejuto Zaera, K. Byczuk, and K. Jabłonowski for the useful discussions. The work was supported by the Foundation for Polish Science through the International Research Agendas Programme co-financed by the European Union within the Smart Growth Operational Programme (Grant No. MAB/2017/1). 
and by Narodowe Centrum Nauki (NCN, National Science Centre, Poland) Project No. 2019/34/E/ST3/00404. 
\end{acknowledgments}

\bibliography{biblio}

\end{document}